\documentclass{emulateapj}
\usepackage{graphics}
\usepackage{natbib}
\bibpunct{(}{)}{;}{a}{}{,}
\usepackage{amssymb}

\newcommand{\rh}{\varrho}
\newcommand{\divv}{\,\mbox{div}}
\newcommand{\grad}{\,\mbox{grad}}

\begin{document}

\title{Influence of random fluctuations in the $\Lambda$-effect on
meridional flow and differential rotation}

\author{Matthias Rempel}

\affil{High Altitude Observatory,
       National Center for Atmospheric Research\footnote{The National
       Center for Atmospheric Research is sponsored by the National
       Science Foundation} , 
       P.O. Box 3000, Boulder, Colorado 80307, USA
      }

\email{rempel@hao.ucar.edu}

\shorttitle{Random fluctuations of $\Lambda$-effect}
\shortauthors{M. Rempel}

\begin{abstract}
We present a mean field model based on the approach taken by 
\citet{Rempel:2005}
in order to investigate the influence of stochastic fluctuations in the
Reynolds stresses on meridional flow and differential rotation. The 
stochastic fluctuations found in the
meridional flow pattern directly resemble the stochastic fluctuations of the
Reynolds stresses, while the stochastic fluctuations in the
differential rotation are smaller by almost two orders of magnitude. 
It is further found that the correlation length and time scale of the 
stochastic fluctuations have only a weak influence on meridional flow, but a 
significant influence on the magnitude of variations in the differential 
rotation. We analyze the energy fluxes within the model to estimate time scales
for the replenishment of differential rotation and meridional flow. We find 
that the time scale for the replenishment of differential rotation ($\sim 10$
years) is nearly four orders of magnitude longer than the time scale for the 
replenishment of meridional flow, which explains the differences in the
response to stochastic fluctuations of the Reynolds stress found for both flow
fields.
\end{abstract}


\section{Introduction}
Recently \citet{Rempel:2005} presented a mean field model for solar 
differential rotation and meridional flow, assuming a 
parameterization of all convective scale processes, most importantly
the turbulent angular momentum transport ($\Lambda$-effect) introduced
by \citet{Kitchatinov:Ruediger:1993}. The model presented by 
\citet{Rempel:2005} and also other mean field models for differential rotation 
\citep[see e.g.][]{Kitchatinov:Ruediger:1995,Ruediger:etal:1998,Kueker:Stix:2001} use a time independent parameterization of the turbulent angular
momentum transport leading to stationary solutions describing the mean flows. 
On the other hand, 3D numerical simulations show a significant temporal 
variation of meridional flow and differential rotation due to the fluctuations 
in the convective motions leading to the Reynolds stresses 
\citep{Miesch:etal:2000,Brun:Toomre:2002}.

Observations of differential rotation are usually based on 3 to 4 month
sets of data and therefore can only tell us something about variations on 
time scales longer than this period \citep{Schou:etal:2002}. The $1\sigma$
error-intervals of inversions within the convection zone are typically 
around a few nHz or less than $1\%$ of the rotation rate. The differential 
rotation also shows a systematic cycle variation, the torsional oscillations, 
which have an amplitude of around $1\%$ of the rotation rate. By contrast 
the variability found in observations of the meridional surface flows is much 
larger compared to the average amplitude.
Over time scales of years \citet{Zhao:Kosovichev:2004} found variations of
the meridional surface flow of the order of $10\mbox{ms}^{-1}$ compared to 
a mean flow amplitude of $20\mbox{ms}^{-1}$. While these variations are most
probably also related to the solar cycle, much less is known about short term
variations, which could be related to the turbulent origin of the flow
itself. Comparison of the long term variations found in differential rotation 
and meridional flow indicates that the relative amplitude of the meridional 
flow variability is almost a factor of $100$ larger than the variability of 
the differential rotation.

In this paper we present a mean field model for differential rotation and
meridional flow following
the approach described in \citet{Rempel:2005} to investigate the influence
of stochastic fluctuations in the Reynolds stresses on meridional flow
and differential rotation. The main intention
of this paper is to investigate how much variability can be expected in
the meridional flow pattern, given the strong constraints set by 
helioseismology on the variability of the 3 to 4 month mean of the 
differential rotation. The possible influence of the solar cycle on
both flow patterns will addressed later in a separate paper.

\begin{figure*}
  \resizebox{\hsize}{!}{\includegraphics{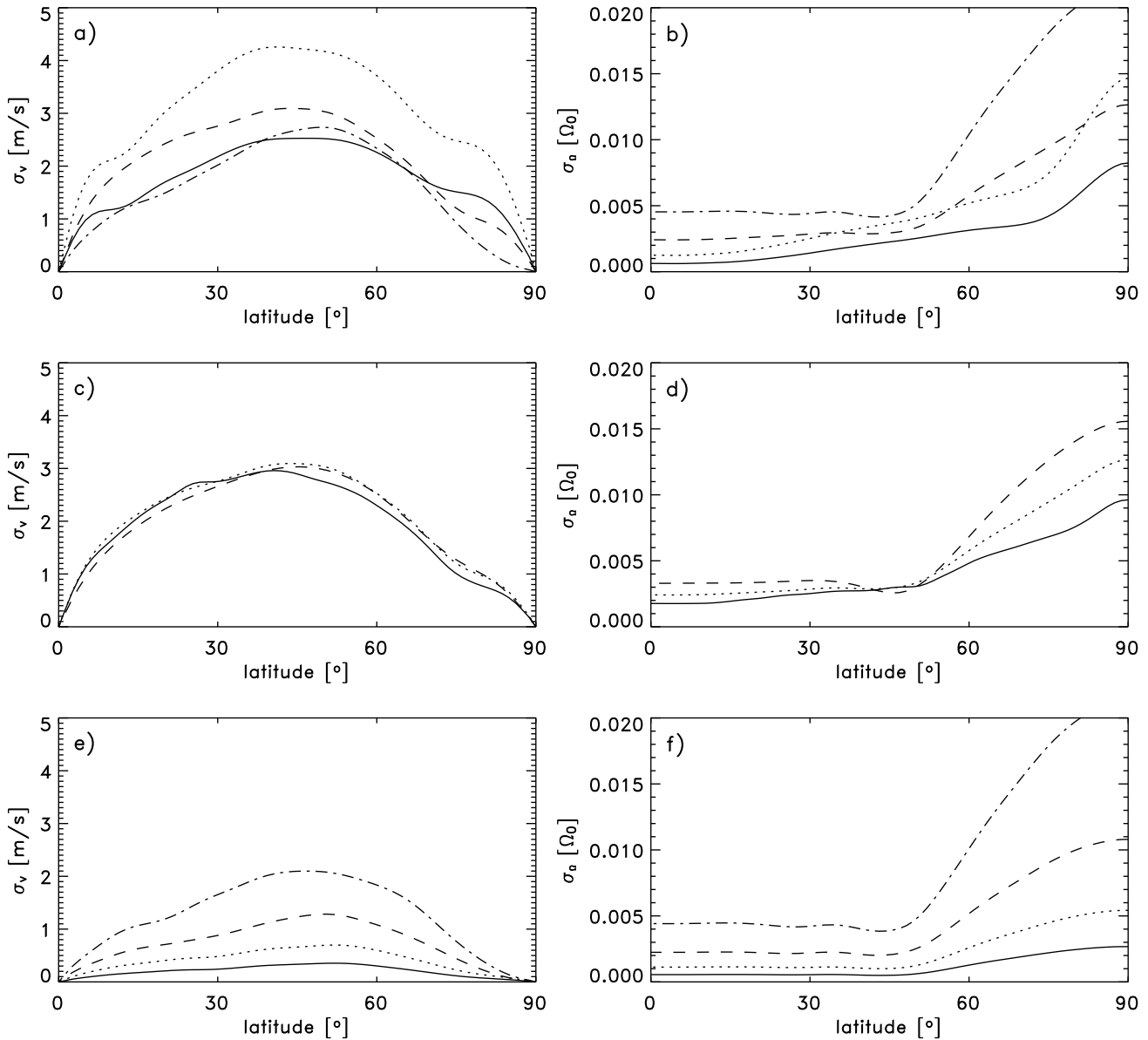}}
  \caption{Variance of meridional flow (left panels) and differential
    rotation (right panels) at the top of the domain $r=0.985 R_{\odot}$. 
    The top panels a) and b) show the total variance of the time series for 
    the cases 1 to 4 (variation of correlation time scale). The solid line
    indicates case 1 ($\tau=0.25\Omega_0^{-1}$), the dotted line case 2
    ($\tau=1\Omega_0^{-1}$), the dashed line case 3 ($\tau=4\Omega_0^{-1}$),
    and dashed-dotted line 4 ($\tau=16\Omega_0^{-1}$). The middle panels 
    c) and d) show cases 5, 3, and 6 (variation of correlation length
    scale). The solid line indicates case 5 ($\Delta=0.05$), the dotted line
    case 3 ($\Delta=0.1$), and the dashed line case 6 ($\Delta=0.2$).
    The bottom panels e) and f) are similar to the top panels, however the
    variances are computed from a four month mean. 
  }
  \label{f1}
\end{figure*}

\section{Model}
For this investigation we use the model described in detail in 
\citet{Rempel:2005}. We add random noise with defined correlation
length and time scales to the parameterization of the turbulent 
Reynolds stress responsible for driving the differential rotation and
meridional flow [see \citet{Rempel:2005}, equations (31) and (32)]:
\begin{eqnarray}
  \Lambda_{r\phi}&=&\Lambda_{\phi r} =
     +{L}(r,\theta)\,\cos(\theta+\lambda(r,\theta))\nonumber\\
     &&(1+c\;\zeta_r(r,\theta)/\sigma_r)  \label{lambda_a}\\
  \Lambda_{\theta\phi}&=&\Lambda_{\phi\theta} =
     -{L}(r,\theta)\,\sin(\theta+\lambda(r,\theta))\nonumber\\
     &&(1+c\;\zeta_{\theta}(r,\theta)/\sigma_{\theta}) \label{lambda_b}
\end{eqnarray}
$L(r,\theta)$ denotes the mean amplitude of the turbulent angular 
momentum flux, whereas $\lambda(r,\theta)$ describes the mean inclination 
of the flux vector with respect to the axis of rotation. 
$L(r,\theta)$ and $\lambda(r,\theta)$ need to be antisymmetric 
across the equator to fulfill the symmetry constraints of the 
$\Lambda$-effect. For further details, see \citet{Rempel:2005}. 

$\zeta_r$ and $\zeta_{\theta}$ denote random functions, which 
will be described below in more detail. The parameter $c$ determines the 
amplitude of the random noise in units of the standard deviations 
$\sigma_r$ and $\sigma_{\theta}$.
By perturbing both components of the turbulent angular momentum flux with
uncorrelated random functions we allow for a change of amplitude and direction
of the angular momentum flux.

In order to generate a random function with a defined correlation length scale,
we construct a 2D random field in the $r$-$\theta$-plane by superposing
Gaussian functions with a fixed width that determines the length scale.
The position in the $r$-$\theta$-plane of individual Gaussian as well as 
the amplitudes are random. We introduce 
a correlation time scale into the problem by performing a temporal
average over the 2D random fields. In principle this averaging could be done
by creating a long time series of a 2D random field and using then a running
mean. However, this approach would require a fairly large amount of memory.
Instead we produce a 2D random field $g$ each time step and solve 
the following equation to introduce a correlations time scale:
\begin{equation}
  \frac{\partial\zeta}{\partial t}=-\frac{\zeta}{\tau_c}
  +\frac{g}{\tau_c}\;,
\end{equation}  
where $\zeta$ denotes the random field with a correlation time scale 
$\tau_c$. The main disadvantage of this approach is that the temporal average 
is always dominated by the most recent realizations of $g$. This 
effect can be reduced by applying this method recursively as follows:
\begin{eqnarray}
  \frac{\partial \zeta_1}{\partial t}&=&-\frac{\zeta_1}{\tau_c/n}
  +\frac{\zeta_2}{\tau_c/n}\nonumber \\
  \frac{\partial \zeta_2}{\partial t}&=&-\frac{\zeta_2}{\tau_c/n}
  +\frac{\zeta_3}{\tau_c/n}\nonumber \\
  &&\ldots\nonumber \\
  &&\ldots\nonumber \\
  \frac{\partial \zeta_n}{\partial t}&=&-\frac{\zeta_n}{\tau_c/n}
  +\frac{g}{\tau_c/n}\;,
\end{eqnarray}
where $\zeta_1$ is the random field used for the simulation. The 
functions $\zeta_2$ to $\zeta_n$ are auxiliary functions required
for the averaging process. The results 
presented in this paper are obtained using a value of $n=5$ for the temporal 
averaging. The free parameters in this approach for generating a 2D
random field are the correlation length scale $\Delta$ and the correlation 
time scale $\tau_c$. $\Delta$ measures the half width of the Gaussians at half
maximum in units of the domain size in the radial and 
latitudinal direction (therefore radial and latitudinal width are not the 
same). For the results presented here we use an amplitude [defined by the 
parameter $c$ in equations (\ref{lambda_a}) and (\ref{lambda_b})] of $0.33$, 
meaning the stochastic fluctuation imposed on the Reynolds stress has a 
$1\sigma$ fluctuation corresponding to $33\%$ of the mean.

In order to allow for a detailed statistical analysis of meridional flow and
differential rotation, we compute during the simulation averages of all 
variables and the square of all variables over the output sampling period,
which allows computing the mean flows and the standard deviations including 
all simulation time steps and therefore avoiding the influence of the output 
sampling.

\section{Results}

\begin{figure*}
  \resizebox{\hsize}{!}{\includegraphics{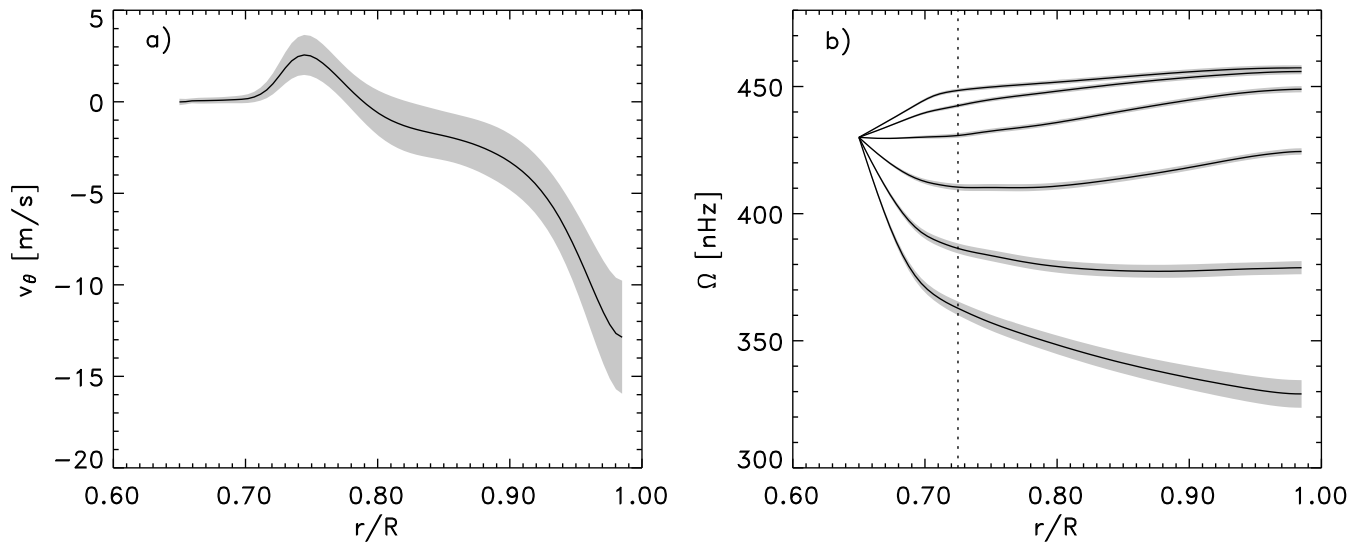}}
  \caption{Depth variation of the standard variations of meridional flow
    and differential rotation for case 3. 
    Panel a) shows the meridional flow profile
    at $45\degr$ latitude and the $1\sigma$ intervals as grey shade.
    Panel b) shows the differential rotation at $0\degr$, $15\degr$, 
    $30\degr$, $45\degr$, $60\degr$, and $90\degr$ latitude with $1\sigma$ 
    intervals as grey shade. 
  }
  \label{f2}
\end{figure*}

\begin{deluxetable}{ccc}
  \tablecaption{Parameters for the stochastic perturbations\label{tab1}}
  \tablehead{ \colhead{case} & \colhead{$\tau_c [\Omega_0^{-1}]$} &
    \colhead{$\Delta$} }
  \startdata
  1 & $0.25$ & $0.1$  \\
  2 & $1.0$  & $0.1$  \\
  3 & $4.0$  & $0.1$  \\
  4 & $16.0$ & $0.1$  \\
  5 & $4.0$  & $0.05$ \\
  6 & $4.0$  & $0.2$   
  \enddata
  \tablecomments{Models 1 to 4 cover a range in the 
    correlation time scale by a factor of $64$, whereas models 5, 3, and 6
    cover correlation lengths scales varying by a factor of $4$. 
    $\Omega_0$ denotes the rotation rate of the solar interior.
  }
\end{deluxetable}

For evaluating the effect of random noise in the turbulent angular momentum
transport we use the model described in the case 1 of \citet{Rempel:2005}
as reference. Table \ref{tab1} summarizes the parameters we use for the 
generation of the random noise fields used to randomize the Reynolds stress
according to equations (\ref{lambda_a}) and (\ref{lambda_b}). All other model 
parameters are given in \citet{Rempel:2005}. The cases 1 to 4
cover a range in the correlation time scale by a factor of $64$ (from roughly
one day to two months), while the series of cases 5, 3, and 6 encompasses a 
variation in the correlation length scale of a factor of $4$ (from a radial 
half width of about 10 Mm to 40 Mm). 

In the following discussion we will focus on the variance of the meridional 
flow $\sigma_v$ and the variance of the differential rotation 
$\sigma_{\Omega}$. Since helioseismic measurements of differential rotation
typically involve 4 month averages, we also compute the variance of 4 month
averages of our model output.

Figure \ref{f1} shows $\sigma_v$ (left column) and $\sigma_{\Omega}$ 
(right column) computed at the top of the domain ($r=0.985R_{\odot}$) 
for the models 1 to 6 listed in Table \ref{tab1}. The top panels show
the total variability for the cases 1 to 4 (change of correlation
time scale), while the middle panels show the cases 5, 3, and 6
(variation of correlation length scale). The bottom panels show the
same properties as the top panels, but computed from a four month
mean. 

The most striking feature visible in panels a) to d) is that while the 
fluctuations of the meridional flow show no systematic variation with 
correlation time and length, the fluctuations of differential rotation show 
a systematic increase with both.
An exception is case 2 in panel a) and b), since $\tau_c=\Omega_0^{-1}$ 
is very close to the time scale of inertial oscillations and therefore
$\sigma_v$ is enhanced because of a resonance. The amplitude of $\sigma_v$
is about $20\%$ to $25\%$ of the mean meridional flow (in our model the 
meridional flow reaches a peak value of around $13.5\,\mbox{m}\,\mbox{s}^{-1}$
at $45\degr$ latitude), which is close to
the noise level added to the Reynolds stress ($33\%$). On the other hand,
except for high latitudes the value of $\sigma_{\Omega}$ is less than
$1\%$ of the reference rotation rate in all cases.

Panels e) and f) show the similar quantities as panels a) and b), however
computed from a four month mean of the time series, similar to the averaging
interval typically used in GONG/MDI inversions of differential rotation
profiles. It is not surprising that the amplitude of the fluctuations 
decreases,
however the effect is stronger in case of the meridional flow, e.g. the 
averaging decreases $\sigma_v$ for case 3 (dashed line) by a factor of $2$,
while $\sigma_{\Omega}$ is nearly unaffected.
  
So far we focused on the variability of meridional flow and differential 
rotation at the top of the domain. Figure \ref{f2} shows the depth 
dependence for the meridional flow at $45\degr$ and the for differential 
rotation at the selected latitudes $0\degr$, $15\degr$, $30\degr$, $45\degr$, 
$60\degr$, and $90\degr$. In both cases solid lines indicate the mean and
the grey shade the $1\sigma$ interval. Shown is case 3 with a correlation
time scale of $4\Omega_0^{-1}$. Both $\sigma_{v}$ and $\sigma_{\Omega}$ show
a maximum at the top of the domain and a nearly monotonic decrease towards
the base of the convection zone. We emphasize that this result
was obtained by using a constant correlation time and length scale 
throughout the convection zone. In a more realistic description it can be
expected that both
correlation time and length scale decrease towards the solar surface, which
according to the trends found in Figure \ref{f1} would lead to
a decrease of $\sigma_{\Omega}$ close to the surface, while $\sigma_{v}$
remains more or less unchanged (except for the enhancement of the signal close
to the resonance). In our investigation we also use a constant amplitude of the
stochastic fluctuations of the Reynolds stress throughout the convection
zone. A depth dependence of this amplitude would influence both, $\sigma_v$
and $\sigma_{\Omega}$ in a similar way.

We computed all models shown in this paper with an amplitude of $c=0.33$ for
the random noise in the $\Lambda$-effect.
Note that $\sigma_v$ and $\sigma_{\Omega}$ scale linearly
with $c$, meaning that the results given here can be rescaled to
apply to a different amplitude of the random noise.

\begin{figure*}
  \resizebox{\hsize}{!}{\includegraphics{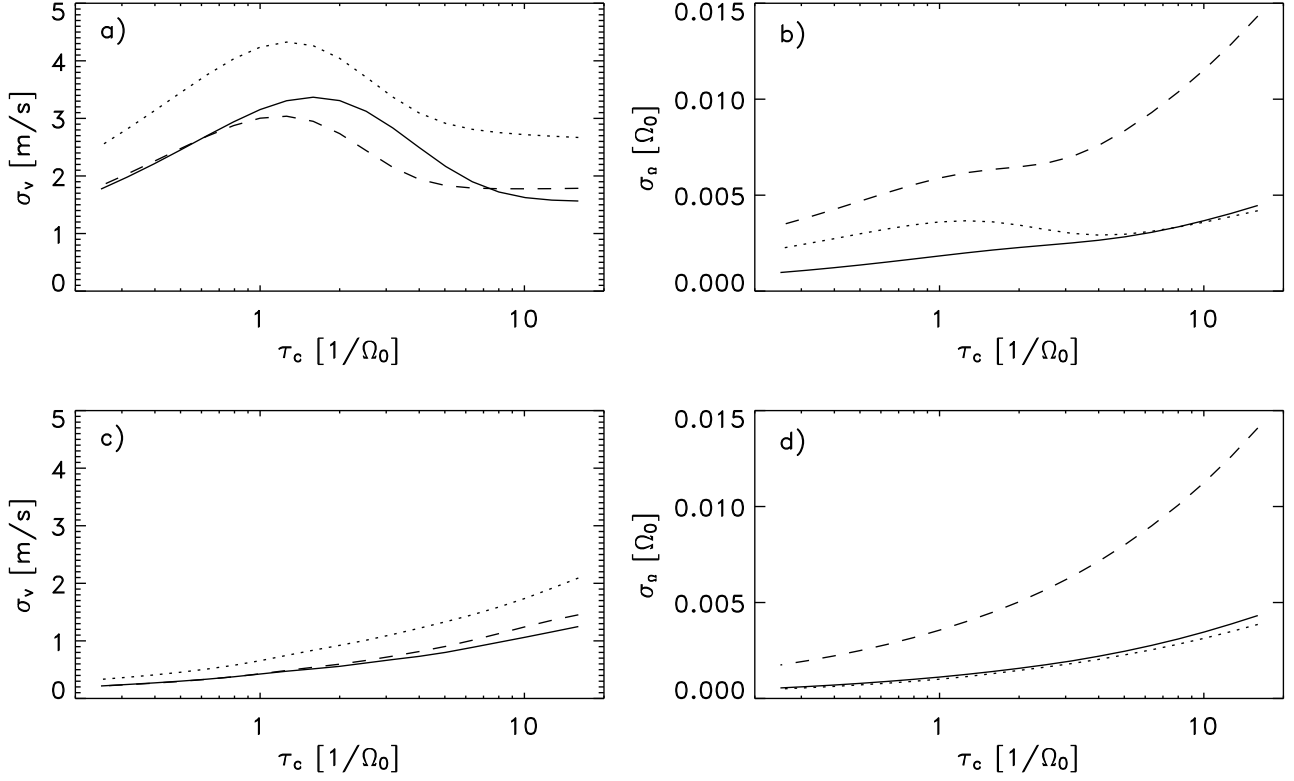}}
  \caption{Variation of the standard variations of meridional flow
    and differential rotation with correlation time scale $\tau_c$ of the 
    random forcing. Panels (a - b) show the total variance of the time series
    as function of $\tau_c$ at the top of the domain for the latitudes
    $22.5\degr$ (solid), $45\degr$ (dotted), and $67.5\degr$ (dashed).
    Panels (c - d) show the variance of the 4 month mean for the same 
    variables. The enhancement of the variance caused by inertia oscillations 
    is clearly visible in panels (a, b), but not in the four month mean shown 
    in panels (c, d).
  }
  \label{f3}
\end{figure*}

We indicated already above that the variance of the meridional flow and
differential rotation is enhanced around $\tau_c\sim\Omega_0^{-1}$ because of
the excitation of inertia oscillations. Figure \ref{f3} shows the
variances of meridional flow and differential rotation at the top of the 
domain ($r=0.985\,R_{\odot}$) for the latitudes $22.5\degr$ (solid), 
$45\degr$ (dotted), and $67.5\degr$ (dashed) as function of $\tau_c$. Panel a) 
shows the total variance of the meridional flow, clearly indicating a 
resonance around $\tau_c\sim 1 - 2\,\Omega_0^{-1}$. Since our time scale 
$\tau_c$ is defined as an
e-folding time scale for the random perturbations, we expect this resonance
around $\tau_c\approx \tau_i/2=\pi/\omega_i$, with the frequency of inertia 
oscillations given by $\omega_i=2\Omega_0$.
This leads to a locations of the resonance around 
\begin{equation}
  \tau_r \approx\frac{\pi}{2\Omega_0}\;.
\end{equation}
Figure \ref{f3}a shows a weak tendency of increasing $\tau_r$ with decreasing 
latitude, which indicates that the geometry of the problem and the 
stratification leads to some confinement of motions in latitude (for
motions in latitude only we would have $\omega_i=2\Omega_0\cos\theta$). 

The resonance also affects the variance of
the differential rotation; however, this influence is much weaker and
disappears at low latitudes (variations in the meridional flow have less 
influence on $\Omega$ close to the equator). The resonance is not visible 
in the variance of the four month mean flows (panels c, d), which shows the 
monotonic increase of the $\sigma_v$ and $\sigma_{\Omega}$ with $\tau_c$ at 
all latitudes as already indicated in Figure \ref{f1}. 

\section{Interpretation}
Our main results presented can be understood in terms
of different response time scales of meridional flow and differential rotation
with respect to changes in the Reynolds stress. The difference in the response
time scale follows from the different size of the kinetic energy 
reservoirs associated with meridional flow and differential rotation.
Here we make a detailed analysis of the energy fluxes within the
differential rotation model in order to estimate typical time scales related
to the meridional flow and differential rotation. In the following derivation
we use the assumptions $\rh_1\ll \rh_0$, $p_1\ll p_0$, and 
$\vert\delta\vert=\vert\nabla-\nabla_{\rm ad}\vert\ll 1$. We also
omit terms proportional to $\divv({\rh_0\mbox{\boldmath$v_m$}})$, which are
very small for the low Mach number meridional flow and vanish for the 
stationary solution finally considered. We start from
equations (2) to (4) of \citet{Rempel:2005} to derive two separate energy 
equations for the meridional flow and differential rotation. To this end we 
multiply equation (2) by $\rh_0 v_r$ and equation (3) by $\rh_0 v_{\theta}$ 
and add both equations. Rearrangement of the different terms leads to:
\begin{eqnarray}
\frac{\partial}{\partial t}&&\left(\rh_0\frac{v_m^2}{2}\right)+
\divv{\mbox{\boldmath$F$}}_m=-(r\sin\theta)^2\rh_0{\mbox{\boldmath$v$}}_m
\cdot\grad\frac{\Omega^2}{2}\nonumber\\
&&-\frac{1}{2}\left[R_{rr}{E}_{rr}+2{R}_{r\theta}
  {E}_{r\theta}+{R}_{\theta\theta}{E}_{\theta\theta}
  +{R}_{\phi\phi}{E}_{\phi\phi}\right]\nonumber\\
&&+v_r\rh_0 g\frac{s_1}{\gamma}\label{e_merid}
\end{eqnarray}
with the energy flux:
\begin{eqnarray}
  {\mbox{\boldmath$F$}}_m&=&\rh_0{\mbox{\boldmath$v$}}_m\left(\frac{v_m^2}{2}
  +\frac{p_1}{\rh_0}
  \right)-\rh_0{\mbox{\boldmath$v$}}_m(r\sin\theta)^2
  \frac{\Omega^2-\Omega_0^2}{2}\nonumber\\
  &&+{\mbox{\bf R }}\cdot{\mbox{\boldmath$v$}}_m
  \label{flux_merid}\;.
\end{eqnarray}
Here {\mbox{\bf R }} denotes the Reynolds stress tensor, 
{\mbox{\bf E}} the deformations tensor. For the definition we refer
to \citet{Rempel:2005}.
Multiplying equation (4) of \citet{Rempel:2005} by 
$\rh_0(r\sin\theta)^2\Omega$ leads to:
\begin{eqnarray}
\frac{\partial}{\partial t}&&\left(\rh_0\frac{(\Omega r\sin\theta)^2}{2}
\right)+\divv{\mbox{\boldmath$F$}}_{\Omega}=\nonumber\\
&&-\nu_t\rh_0(r\sin\theta)^2\left[
  \left(\frac{\partial\Omega}{\partial r}\right)^2+
  \left(\frac{1}{r}\frac{\partial\Omega}{\partial \theta}\right)^2\right]
\nonumber\\
&&-\nu_t\rh_0 r\sin\theta\left(\frac{\partial\Omega}{\partial r}
  \Lambda_{r\phi}+\frac{1}{r}\frac{\partial\Omega}{\partial \theta}
  \Lambda_{\theta\phi}\right)\nonumber\\
&&+(r\sin\theta)^2\rh_0{\mbox{\boldmath$v$}}_m\cdot\grad\frac{\Omega^2}{2}
  \label{e_diff}
\end{eqnarray}
with the energy flux:
\begin{equation}
  {\mbox{\boldmath$F$}}_{\Omega}=\rh_0{\mbox{\boldmath$v$}}_m\left(\Omega r 
  \sin\theta\right)^2
  -r\sin\theta\Omega{\mbox{\boldmath$R$}}_{\phi}\label{flux_diff}\;.
\end{equation}
Here ${\mbox{\boldmath$R$}}_{\phi}$ denotes a vector with components 
$R_{r\phi}$ and $R_{\theta\phi}$.
Since for the boundary conditions we use, the energy fluxes [equations 
(\ref{flux_merid}) and (\ref{flux_diff})] vanish at the boundaries, 
the divergence terms in equations (\ref{e_merid}) and (\ref{e_diff}) do not 
contribute after integration over the entire domain, leading to
the energy balances:
\begin{eqnarray}
  \frac{\partial E_m}{\partial t}&=&Q_{C}+Q^m_{visc}+Q^m_{B}\;,\\
  \frac{\partial E_{\Omega}}{\partial t}&=&-Q_{C}
  +Q^{\Omega}_{visc}+Q^{\Omega}_{\Lambda}\;,
\end{eqnarray}
with the terms
\begin{eqnarray}
  E_m&=&\int\mbox{d}V\rh_0\frac{v_m^2}{2}\;,\nonumber\\
  Q_{C}&=&-\int\mbox{d}V
  (r\sin\theta)^2\rh_0{\mbox{\boldmath$v$}}_m\cdot\grad\frac{\Omega^2}{2}\;,
  \nonumber\\ 
  Q^m_{\nu}&=&-\int\mbox{d}V\frac{1}{2}\left[R_{rr}{E}_{rr}+2{R}_{r\theta}
  {E}_{r\theta}+{R}_{\theta\theta}{E}_{\theta\theta}\right.\nonumber\\
  &&\left.+{R}_{\phi\phi}{E}_{\phi\phi}\right]\;,\nonumber\\
  Q^m_{B}&=&\int\mbox{d}V\,v_r\rh_0 g\frac{s_1}{\gamma}\;,\nonumber\\
  E_{\Omega}&=&\int\mbox{d}V\frac{1}{2}\rh_0(r\sin\theta)^2
  \left(\Omega^2-\Omega_0^2\right)\;,\nonumber\\
  Q^{\Omega}_{\nu}&=&-\int\mbox{d}V\nu_t\rh_0(r\sin\theta)^2\left[
  \left(\frac{\partial\Omega}{\partial r}\right)^2+
  \left(\frac{1}{r}\frac{\partial\Omega}{\partial \theta}\right)^2\right]
  \;,\nonumber\\
  Q^{\Omega}_{\Lambda}&=&-\int\mbox{d}V\nu_t\rh_0 r\sin\theta
  \left(\frac{\partial\Omega}{\partial r}
  \Lambda_{r\phi}+\frac{1}{r}\frac{\partial\Omega}{\partial \theta}
  \Lambda_{\theta\phi}\right)\;.
\end{eqnarray}
Here
\begin{equation}
  \int\mbox{d}V=4\pi\int_{r_{min}}^{r_{max}}\mbox{d}r\int_{0}^{\pi/2}
  \mbox{d}\theta r^2\sin\theta
\end{equation}
denotes the integral over the entire volume of the sphere from $r=r_{min}$
to $r=r_{max}$. We emphasize that we solve our model only for the
northern hemisphere but we compute from that the energy conversion for
the entire sphere. $E_m$ and $E_{\Omega}$ denote energy available in the 
reservoir of meridional flow and differential rotation, respectively. For the 
latter we subtracted the core rotation rate to obtain a value representative
for the differential rotation rather than total rotation. $Q_{C}$ denotes
the amount of energy that is converted by means of the Coriolis force
(which is a sink for differential rotation and source for meridional flow). 
The terms $Q^m_{\nu}$ and $Q^{\Omega}_{\nu}$ represent the losses through 
viscous
dissipation for meridional flow and differential rotation, respectively.
$Q^{\Omega}_{\Lambda}$ is the energy which is converted through the 
$\Lambda$-effect from internal energy to energy of differential rotation,
and $Q^m_{B}$ denotes the work of the meridional flow against the buoyancy
force arising from the entropy perturbation within the convection zone.  

\begin{figure}
  \resizebox{\hsize}{!}{\includegraphics{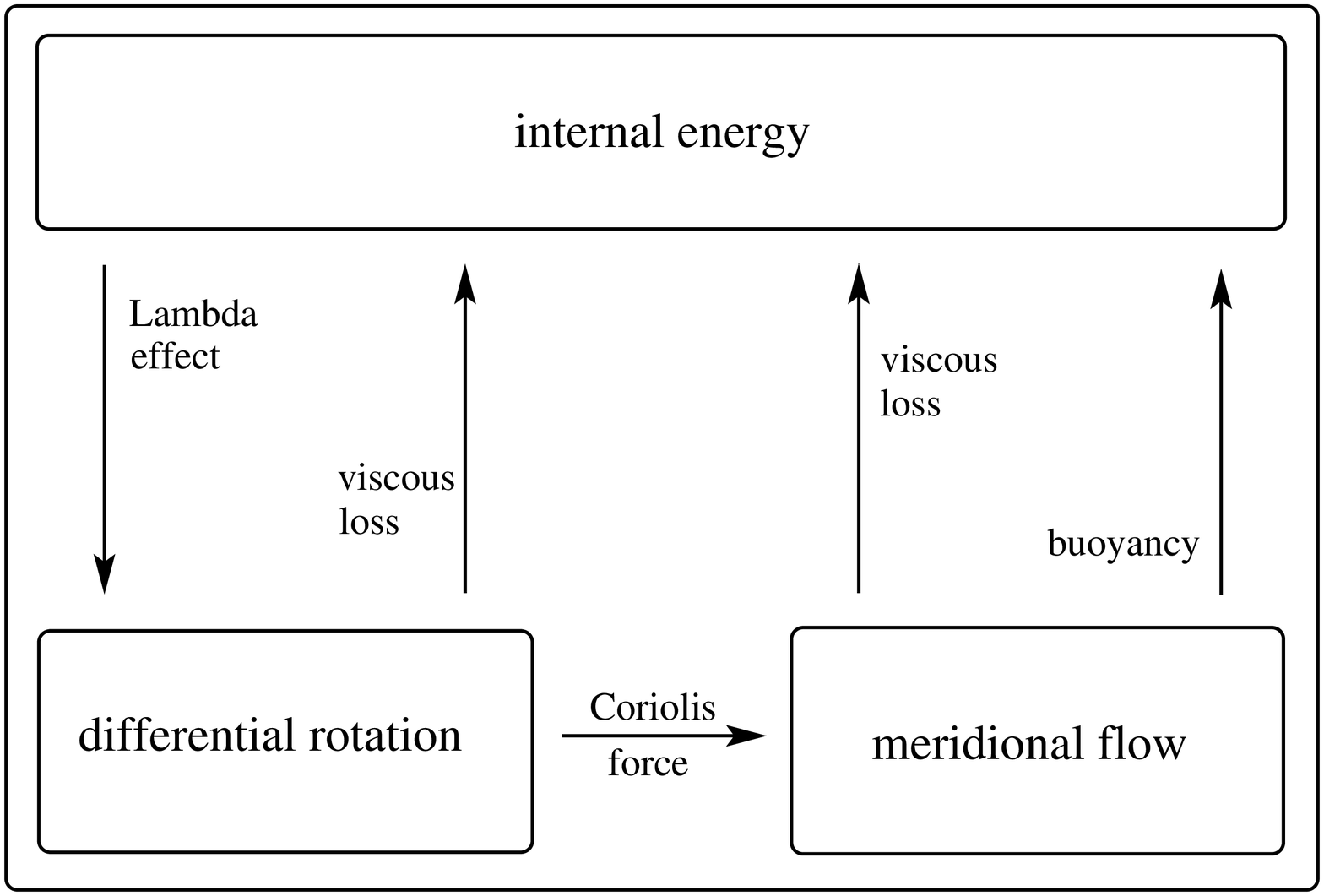}}
  \caption{Schematic view of energy fluxes in differential rotation model.
  }
  \label{f4}
\end{figure}

For the stationary reference model we can write the energy balance as:
\begin{eqnarray}
  Q_{C}&=&-Q^m_{visc}-Q^m_{B}\;,\\
  Q^{\Omega}_{\Lambda}&=&-Q^{\Omega}_{visc}+Q_{C}\;,
\end{eqnarray}
where the left-hand side terms are the sources and the right-hand
terms the sinks for the reservoirs of meridional flow and differential
rotation, respectively. Figure \ref{f4} shows a qualitative diagram indicating
the energy fluxes between the different energy reservoirs. A quantitative 
analysis of the model 1 of \citet{Rempel:2005}, which we used in this paper 
as reference model, leads to
\begin{eqnarray}
  E_{\Omega}&=&1.55\times 10^{33}\,\mbox{J}\;,\nonumber\\
  Q^{\Omega}_{\Lambda}&=&6.6\times 10^{24}\,\mbox{W}=0.017\,L_{\odot}
  \;,\nonumber\\
  \tau_{\Omega}&=&\frac{E_{\Omega}}{Q^{\Omega}_{\Lambda}}=7.5\,\mbox{years}
  \;,\nonumber\\
  Q^{\Omega}_{\nu}&=&-0.54\;Q^{\Omega}_{\Lambda}\;,\nonumber\\
  Q_{C}&=&0.46\;Q^{\Omega}_{\Lambda}\label{energy_conversion_diff}
\end{eqnarray}
for the differential rotation and to
\begin{eqnarray}
  E_m&=&10^{29}\,\mbox{J}\;,\nonumber\\
  Q_{C}&=&3\times 10^{24}\,\mbox{J}=0.008\,L_{\odot}\;,\nonumber\\
  \tau_m&=&\frac{E_m}{Q_{C}}=9.4\,\mbox{hours}\;,\nonumber\\
  Q^{m}_{\nu}&=&-0.02\;Q_{C}\;,\nonumber\\
  Q^{m}_{B}&=&-0.98\;Q_{C}\label{energy_conversion_merid}
\end{eqnarray}
for the meridional flow.
  
In order to maintain the differential rotation in this model, the $\Lambda$
-effect has to convert an amount of energy equivalent to around $1.7\%$ of 
the solar luminosity. Comparing this energy flux to the energy stored in the
reservoir of differential rotation leads to a time scale of around $8$ years
for the replenishment of energy. Around $54\%$ of the energy converted by
the $\Lambda$-effect returns directly to the reservoir of internal energy
through viscous dissipation, while $46\%$ flows through the action of the
Coriolis force into the reservoir of meridional flow. Dividing the energy
in this reservoir by the energy flux related to the Coriolis force leads to a
time scale for the replenishment of energy in the meridional flow of only
$10$ hours, which is around a factor of $6500$ shorter than the corresponding
time scale for the differential rotation. The direct viscous loss does not 
play an important role for the meridional flow (only around $2\%$ of the energy
is dissipated); most of the energy returns to the reservoir of internal
energy through work against the buoyancy force. 

The numbers presented here
vary with different model assumptions made but the fact that $\tau_{\Omega}$
is around 4 orders of magnitude longer than $\tau_m$ is a very robust result, 
since it follows mainly from the different sizes of the energy reservoirs of 
differential rotation and meridional flow. Another very robust result is that 
the viscous dissipation does not play an important
role for the meridional flow (unless the assumed turbulent viscosity would
be more than one order of magnitude larger than the value of
$5\times 10^8\,\mbox{m}^2\,\mbox{s}^{-1}$ used here). This emphasizes the
importance of the pole-equator entropy variation for avoiding the
Taylor-Proudman state, as discussed in \citet{Rempel:2005}, since the work
of the meridional flow against the buoyant force is the primary sink of
energy.  
 
From the very short response time scale of $\tau_m\sim 10$ hours it can
be expected that the meridional flow is affected nearly instantaneously
by fluctuations in the Reynolds stress, while the differential rotation 
only responses to a long term average over several years. As a consequence,
the relative amplitudes of fluctuations in meridional flow and differential 
rotation are expected to scale with a factor of 
$\sim \sqrt{\tau_{\Omega}/\tau_m}\sim 80$, which is seen in the results 
presented before. 

An order of magnitude estimate based on the energy yields:
\begin{eqnarray}
  \Delta E_{m}&\sim& v^{m}_{\rm rms}\Delta v^{m}_{\rm rms}\;,\\
  \Delta E_{\Omega}&\sim& v^{\Omega}_{\rm rms}\Delta v^{\Omega}_{\rm rms}\;.
\end{eqnarray}
Given the fact that the amount of energy transfered into the meridional
flow is roughly half the energy transferred into the differential rotation,
the expected change in the flow amplitudes scales like
\begin{equation}
  \frac{\Delta v^{m}_{\rm rms}}{\Delta v^{\Omega}_{\rm rms}}\sim 
    \frac{1}{2}\frac{v^{\Omega}_{\rm rms}}{v^{m}_{\rm rms}}\sim 100\;,
\end{equation}
(with $v^{\Omega}_{\rm rms}\sim 1\,\mbox{km}\,\mbox{s}^{-1}$ and 
$v^{m}_{\rm rms}\sim 5\,\mbox{m}\,\mbox{s}^{-1}$. 
This is close to the ratio we 
obtained in the more detailed analysis of the energy fluxes within the
model.

Since the large kinetic energy reservoir of the differential rotation tends
to average fluctuations, a systematic variation of $\sigma_{\Omega}$ with
correlation time scale and length scale is seen in the data (longer lasting,
large scale perturbations manifest easier in the differential rotation). On 
the other hand, the
meridional flow shows a nearly immediate response to changes in the
Reynolds stress. Therefore $\sigma_v$ is mainly determined by the amplitude
of the random fluctuations in the Reynolds stress assumed and not by the 
correlation time and length scale (unless the time scale for fluctuations 
in the Reynolds stress is much shorter than $\tau_m$, which is not 
reasonable in the bulk of the convection zone). The only exception is the 
resonance found in case 2, which enhances the amplitude by about $50\%$.

\section{Conclusions} 
We presented a mean field model for differential rotation and meridional flow
including stochastic fluctuations in the Reynolds stress driving the 
differential rotation. We found that the variations in the differential 
rotation are around two orders of magnitude less than the variability observed
in the meridional flow pattern. For example the case 3 discussed above shows a
$1\sigma$ variation of the 4 month 
mean of less than $0.3\%$ of the rotation rate for the differential rotation 
(in latitudes lower than $50\degr$), while the $1\sigma$ variation for the 
meridional flow is around $8\%$ of the mean flow value (4 month mean) or
even $20\%$ of the mean flow value if also the short term variation is 
considered. The maximum amplitude of the short term meridional flow 
variability in this models is around $10\mbox{ms}^{-1}$ and therefore close 
to $100\%$ of the mean flow. 

This result is a consequence of different response time scales of the
differential rotation and meridional flow to changes in the Reynolds stress.
Our model shows that the energy flows through the reservoir of differential 
rotation and meridional flow are comparable, while the kinetic energy of 
differential rotation exceeds that of meridional flow by about 4 orders of 
magnitude. As a consequence the meridional flow shows a nearly
instantaneous response (on a time scale of less than a day) to changes
in the Reynolds stress, while the differential rotation is affected only
by changes that have a long term average of at least $10$ years.

Our result is in agreement with observations, which indicate that the 
$1\sigma$ variation of the 4 month differential rotation inversions is a few 
nHz, but also show that the variations in the meridional flow can be of
up to $10\mbox{ms}^{-1}$. However, most of the meridional flow measurements
so far only address long term variability (time scale of years) rather than
short term variability, which is the main focus of the model presented here.

We conclude that a fairly significant amount of random noise in the Reynolds
stresses driving differential rotation can be tolerated without leading to
variations in the differential rotation contradicting helioseismic inversions.
However, the meridional flow will show significant variability on the same 
order of magnitude as the fluctuations of the Reynolds stress.

We found that the random forcing through the Reynolds stress excites
inertia oscillations for times scales $\tau_c \sim 1 - 2\,\Omega_0^{-1}
\sim 4 - 8\,\mbox{days}$. This resonance enhances the total variance
$\sigma_v$ by about $50\%$ and has a visible but much weaker effect
on $\sigma_{\Omega}$. However, this effect is not visible if the
variance of the four month mean is considered. Therefore this effect 
could be observable in the variability of daily flow maps of the 
surface meridional flow but not in helioseismic inversions normally
considering longer averages.

\acknowledgements
The author thanks P.~A. Gilman for stimulating discussions and helpful
comments on a draft of this manuscript and the anonymous referee for a
helpful review.

\bibliographystyle{natbib/apj}
\bibliography{natbib/apj-jour,natbib/papref}

\end{document}